\documentclass[twocolumn]{aastex62}
\usepackage{amsmath}

\usepackage{float}
\graphicspath{{./}{figures/}}

\shorttitle{Intelligent Responses to Our Technological Signals Will Not Arrive In Fewer Than Three Millennia}
\shortauthors{Siraj \& Loeb}

\begin{document}
\title{Intelligent Responses to Our Technological Signals Will Not Arrive In Fewer Than Three Millennia}

\email{amir.siraj@cfa.harvard.edu, aloeb@cfa.harvard.edu}

\author{Amir Siraj}
\affil{Department of Astronomy, Harvard University, 60 Garden Street, Cambridge, MA 02138, USA}

\author{Abraham Loeb}
\affiliation{Department of Astronomy, Harvard University, 60 Garden Street, Cambridge, MA 02138, USA}




\begin{abstract}

What is the chance we start a conversation with another civilization like our own? Our technological society produced signals that could be received by other extraterrestrial civilizations, within a sphere around us with a radius of $\sim 10^2$ light years.  Given that, the Copernican principle provides a lower limit on the response time that we should expect from transmitters on Earth-like planets around Sun-like stars. If our civilization lives longer, the expected number of responses could increase. We explore the chance of detecting a response in the future, and show that a response should only be expected to arrive after a few millennia.

\end{abstract}

\keywords{SETI}


\section{Introduction}

The search for extraterrestrial intelligence (SETI) can be divided into two broad categories: searches for generic signals, and searches for signals intended specifically for Earth \citep{LingamLoeb21}. Here, we consider the latter category, specifically the case of an ETI responding to a technosignature from Earth, which we dub the search for extraterrestrial responding intelligence (SETRI). In other words, we focus on humanity `starting a conversation' with an ETI. Radio waves are the most obvious and longstanding example of a techonisgnature produced on Earth, dating back approximately 132 years. For reviews of radio SETI, see \cite{2020scia.book....9D} and \cite{LingamLoeb21}.

The simplest assumption one can make about the emergence of life on Earth is that it represents a single outcome of a random process within the context of all Earth-like planets orbiting Sun-like stars. This contrasts with the pessimistic `rare Earth' hypothesis \citep{2000rewc.book.....W}, and represents an application of the Copernican principle, which asserts that we are not privileged observers of the Universe. Previous works that have applied the Copernican principle to SETI include \cite{1993Natur.363..315G}, \cite{2012PNAS..109..395S}, \cite{2019AsBio..19...28L}, \cite{2020ApJ...896...58W}, and \cite{2021arXiv210104118S}.

Under the Copernican principle, \cite{2021arXiv210104118S} presented a derivation of the likelihood that a technological civilization presently exists around any given Sun-like star, given the measured abundance of Earth-like planets around Sun-like stars, the timing of the origin of life on Earth relative to the Sun's lifetime, and the period of time over which radio communication has existed on Earth. This was paired with a uniform prior distribution, since it is the prior that adopts no artificial limit truncations as well as representing the most optimistic choice, since it is biased towards values on the scale of unity. \cite{2021arXiv210104118S} showed that \textit{Breakthrough Listen Candidate 1} is very unlikely to be a possible technosignature originating from the Alpha Centauri system on this basis alone.

We ask the following question: is it plausible to expect a response from a conversation we started? The most optimistic assumption possible is that all ETIs are ETRIs, so here we adopt the limiting case derivation of the likelihood that an ETI presently exists around any given Sun-like star as a proxy for the likelihood of an ETRI existing, combined with the horizon out to which such communication could take place given the relatively recent emergence of radio technology on Earth, to demonstrate that it is not plausible to receive an intelligent response to our own radio signals \textit{at present}. We explore the time dependency of this question, since both the inferred distribution of lifetimes for technological civilizations and the horizon out to which signals can travel increase with time. We derive the timescales out to which humanity should expect no responses to technological signals from civilizations on Earth-like planets orbiting Sun-like stars, even under the most optimistic assumptions possible in a statistical framework.

In Section \ref{sec:f}, we introduce the methods used here. In Section \ref{sec:rs} we present our results. Finally, in Section \ref{sec:d} we discuss key implications. 


\section{Methods}
\label{sec:f}

We adopt the formalism in \cite{2021arXiv210104118S}, in which technosignatures on Earth-like planets orbiting Sun-like stars have some mean emergence timescale, $\tau_{begin}^{\mu}$, and some mean extinction timescale, $\tau_{end}^{\mu}$, where $\tau_{end}^{\mu} > \tau_{begin}^{\mu}$, $\tau_{end}^{\mu} - \tau_{begin}^{\mu} = \tau_{obs}^{\mu}$, and $\tau_{obs}^{\mu} \ll \tau_{begin}^{\mu} \sim \tau_{end}^{\mu}$.

The upper limit on the probability that a single Sun-like star presently hosts a technological civilization is, 

\begin{equation}
\frac{N_{response}}{N_{\star}} = \frac{n(\{\tau_{begin} \; \vert \; \tau_{begin} < \tau_{\star}\})}{n(\{\tau_{begin}\})} \; \cdot \; \frac{\tau_{obs}^{\mu}}{ \tau_{\star}} \; \cdot \; n_{EL} \; \; ,
\end{equation}

where $\tau_{\star} \sim 5.5 \mathrm{\; Gyr}$ is the timescale on which Earth-like planets in the habitable zones of Sun-like stars remain unaffected by inevitable runaway greenhouse heating due to the expansion of the parent star \citep{2008MNRAS.386..155S}, and where $n_{EL}$ is the number of Earth-like planets per Sun-like star, $N_{response}$ is the upper limit on the number of responses we could expect to receive to our technological communications and $N_{\star}$ is the number of Sun-like stars that could communicate with Earth within $\tau_{obs}^{\mu}$.

The horizon out to which a civilization hosted by a Sun-like star could communicate with the Earth within time $\tau_{obs}$, assuming that the outgoing signal is electromagnetic (traveling at $c$) and the incoming signal travels at speed $v$ ($v = c$ for electromagnetic signals, $v < c$ for physical probes), is $R = \tau_{obs} c v / (c + v)$. We ignore the Sun's motion the short time period $\tau_{obs}$.

For $R \leq 100 \mathrm{\; pc}$, $N_{\star} = (4\pi/3) \; R^3 \; n_{\star}$, and for $R > 100 \mathrm{\; pc}$, $N_{\star} = (4\pi/3)\; (100 \mathrm{\; pc})^3 \; (R / 100 \mathrm{\; pc})^2 \; n_{\star}$, where the local number density of Sun-like (G) stars $n_{\star} \sim 3 \times 10^{-3} \; \mathrm{pc^{-3}}$ and $100 \mathrm{\; pc}$ is the Galactic scale height for stars with age of order, $\tau_{\star} \sim 5.5 \mathrm{\; Gyr}$ \citep{2017MNRAS.470.1360B}.

Since the existence of technological civilizations around Sun-like stars should be independent from each other, the most optimistic assumption possible about the underlying processes for emergence and extinction can be expressed as a Poisson process with a uniform prior, which favors probabilities of order unity. The overall expression for the upper limit on the number of ETRIs on Earth-like planets orbiting Sun-like stars is,
\begin{equation}
    N_{response} = (\tau_{obs}^{\mu} / \tau_{\star}) \cdot (1 - e^{-(\tau_{\star} / \tau_{begin}^{\mu})}) \cdot n_{EL} \cdot N_{\star} \; \; ,
\end{equation}
where $\tau_{obs}^{\mu}$, $\tau_{\star}$, and $\tau_{begin}^{\mu}$ are stochastically determined as described in \cite{2021arXiv210104118S}, and $N_{\star}$ is computed as outlined above. Note that for times in the future, $\tau_{obs} > 132 \mathrm{\; yrs}$, the transformation in the probability distribution for $\tau_{obs}^{\mu}$ is linearly related to the change in $\tau_{obs}$. On the other hand, the dependence of $N_{\star}$ on $\tau_{obs}$ is non-linear; cubic out to $R \sim 100 \mathrm{ \; pc}$ and quadratic farther out.

To produce an estimate, the upper limit $N_{response}$ should be multiplied by all the relevant factors in the Drake equation (for a review, see \citealt{2015dreq.book.....V}). 

\section{Results}
\label{sec:rs}
\begin{figure}
 \centering
\includegraphics[width=1\linewidth]{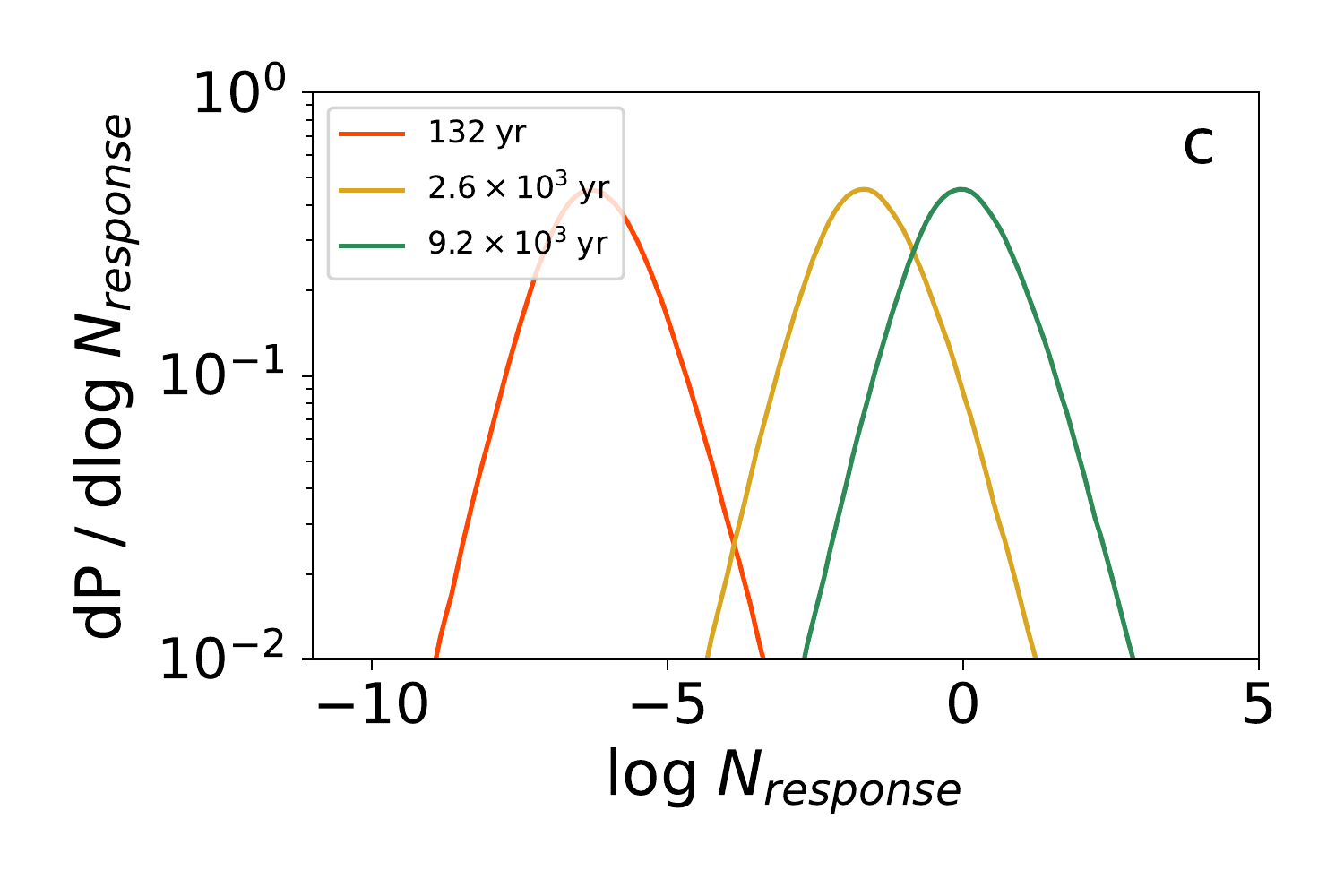}

\caption{Probability distributions of the upper limits on the number of responses we could expect to our receive to our technological communications, for a response signal speed of $c$. Shown here are $\mathrm{dP} / \mathrm{d}\log{N_{response}}$ at three times: $\tau_{obs} = 132 \mathrm{\; yr}$ (present), $\tau_{obs} = 2.6 \times 10^3 \mathrm{\; yr}$ (median of $N_{response}$ of order unity), and $\tau_{obs} = 9.2 \times 10^3 \mathrm{\; yr}$ (95\% upper limit of $N_{response}$ of order unity).}
\label{fig:c}
\end{figure}

We first consider electromagnetic (including radio) responses to our radio signals. Since we have produced radio signals that could be received by other ETIs, within a sphere around us with a radius of 132 light years, the location of ETRIs at present day is limited to a sphere of radius 66 light years. We apply the methods described in Section \ref{sec:f} to derive the probability distribution for $N_{response}$, the upper limit on the number of responses we could expect to receive to our technological communications from civilizations on Earth-like planets orbiting Sun-like stars at the present time, as well as for the 50 and 95 percentile values for $N_{response}$ being of order unity. The latter corresponds to a conservative estimate on the timescale over which we should not expect responses of any sort to our technological signals, since $N_{response}$ is already an upper limit. We find the $95\%$ lower limit for $\tau_{obs}$ to be $2.6 \times 10^3 \; \mathrm{yr}$, and that the median of the upper limit $N_{response} \sim 1$ for $\tau_{obs} \sim 9.2 \times 10^3 \mathrm{\; yr}$. Figure \ref{fig:c} displays the corresponding probability distributions.

\begin{figure}
 \centering
\includegraphics[width=1\linewidth]{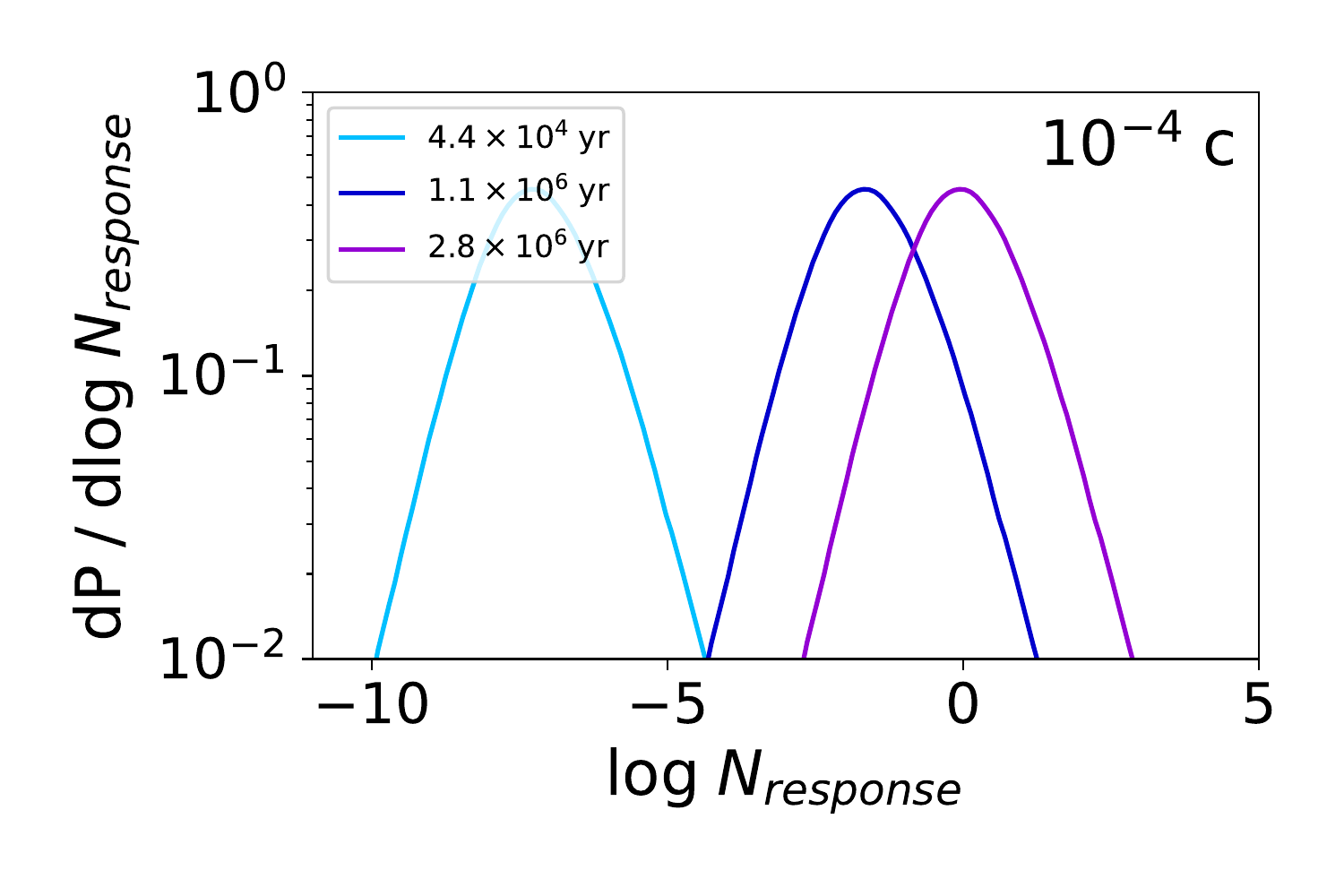}

\caption{Probability distributions of the upper limits on the number of responses we could expect to our receive to our technological communications, for a response signal speed of $10^{-4} c$. Shown here are $\mathrm{dP} / \mathrm{d}\log{N_{response}}$ at three times: $\tau_{obs} = 4.4 \times 10^4 \mathrm{\; yr}$ (travel time to the nearest star), $\tau_{obs} = 1.1 \times 10^6 \mathrm{\; yr}$ (median of $N_{response}$ of order unity), and $\tau_{obs} = 2.8 \times 10^6 \mathrm{\; yr}$ (95\% upper limit of $N_{response}$ of order unity).}
\label{fig:1em4c}
\end{figure}

\begin{figure}
 \centering
\includegraphics[width=1\linewidth]{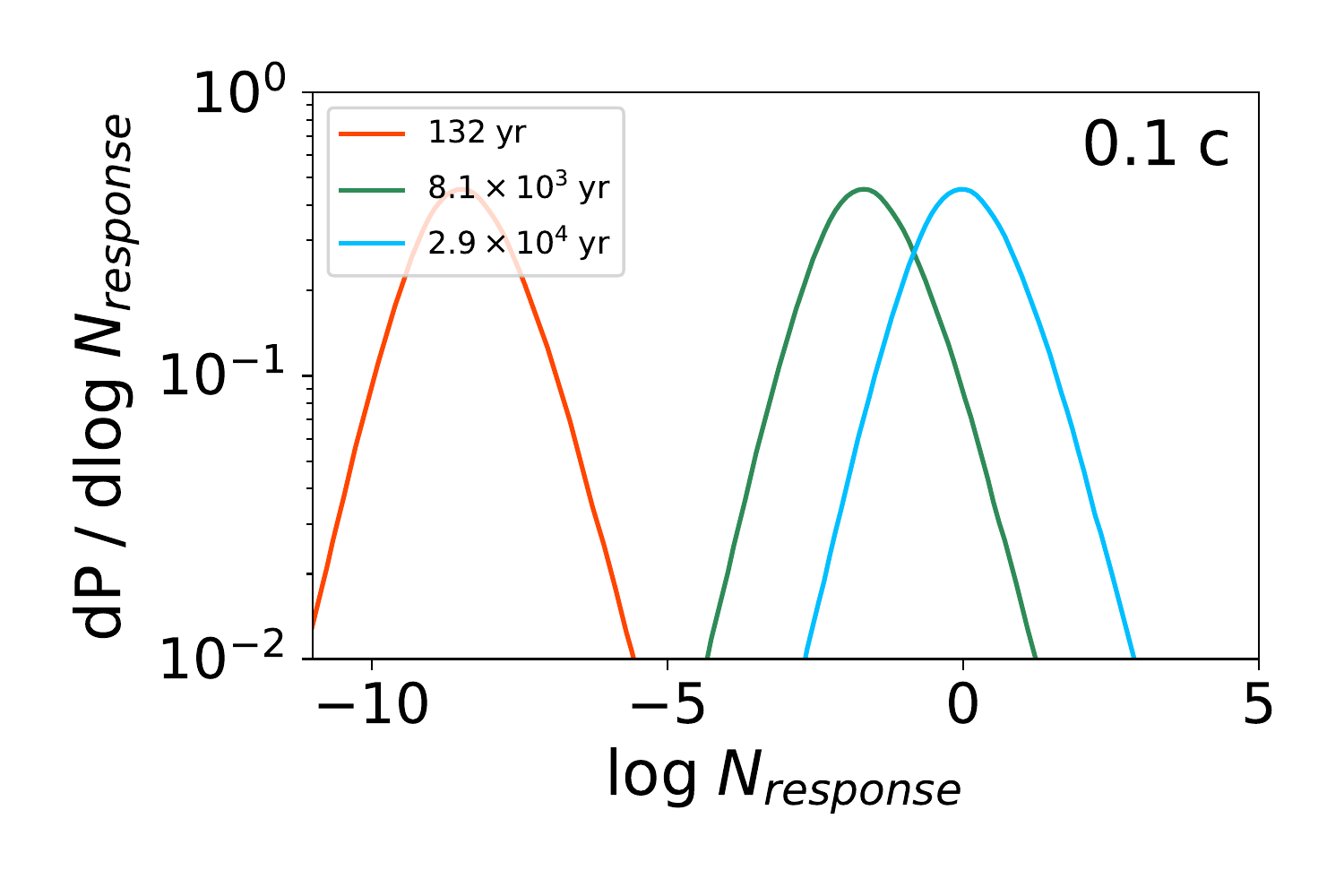}

\caption{Probability distributions of the upper limits on the number of responses we could expect to our receive to our technological communications, for a response signal speed of $0.1 c$. Shown here are $\mathrm{dP} / \mathrm{d}\log{N_{response}}$ at three times: $\tau_{obs} = 132 \mathrm{\; yr}$ (present), $\tau_{obs} = 8.1 \times 10^3 \mathrm{\; yr}$ (median of $N_{response}$ of order unity), and $\tau_{obs} = 2.9 \times 10^4 \mathrm{\; yr}$ (95\% upper limit of $N_{response}$ of order unity).}
\label{fig:1em1c}
\end{figure}

\begin{figure}
 \centering
\includegraphics[width=1\linewidth]{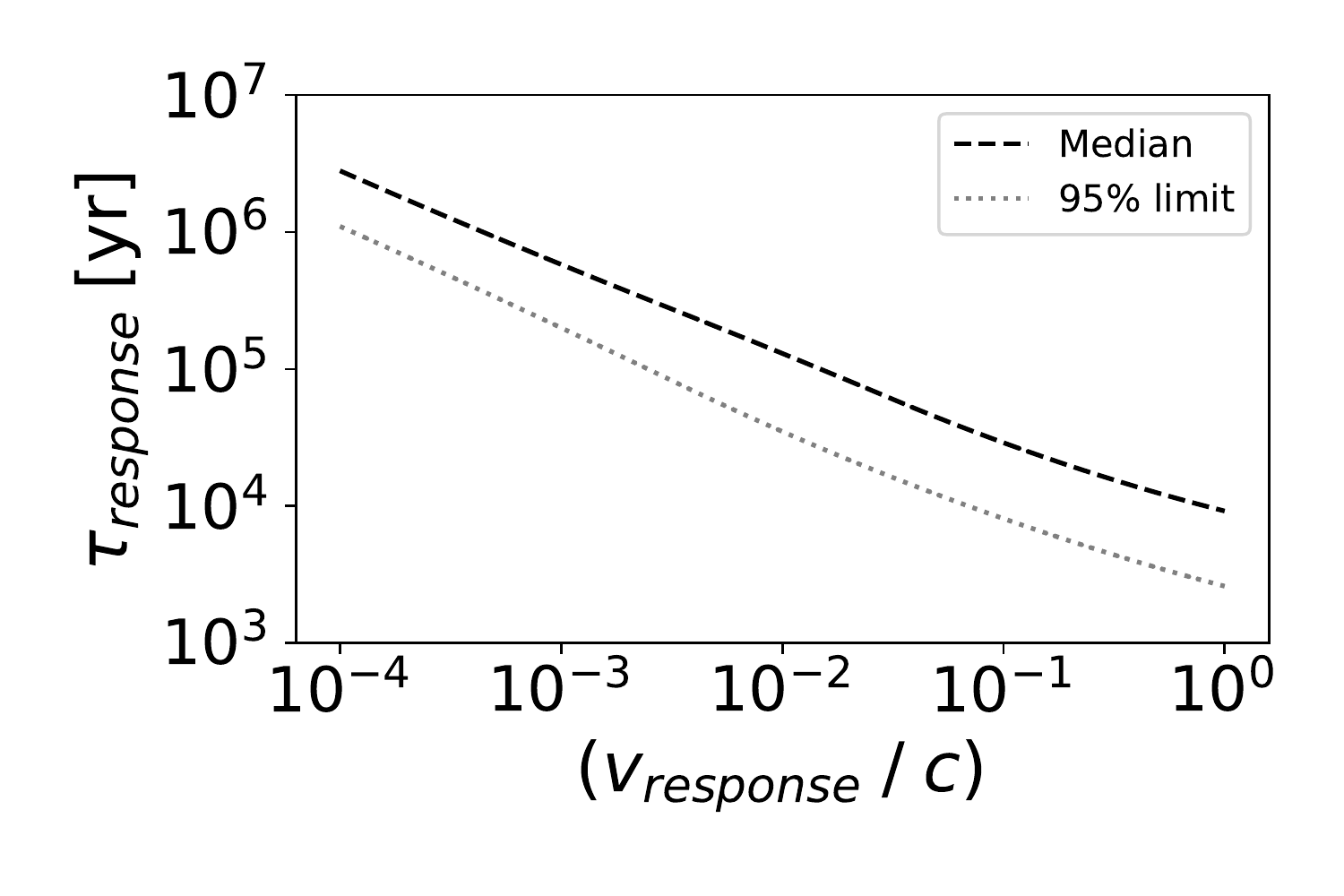}

\caption{The timescale over which responses to our technological communication will not arrive, as a function of response signal speed ($c$ for electromagnetic responses, $10^{-4} \; c$ for rockets with chemical propulsion). Show here are the values of $\tau_{response}$ such that the 50 and 95 percentiles values for $N_{response}$ are of order unity, respectively.}
\label{fig:ts}
\end{figure}

We next consider physical probe responses to our radio signals. The number of ETRIs from which we could receive such responses is dependent on the propulsion speed, and is necessarily smaller than the number of ETRIs we could receive electromagnetic responses from. Applying the methods described in Section \ref{sec:f}, we find that, for chemical rockets with $v \sim 30 \mathrm{\; km \; s^{-1}} = 10^{-4} \; c$ the median of $N_{response}$ is of order unity for $\tau_{obs} = 2.8 \times 10^6 \mathrm{\; yr}$ and the 95\% upper limit of $N_{response}$ is of order unity for $1.1 \times 10^6 \mathrm{\; yr}$, implying that we should not expect a response from a chemically propelled probe before then (probability distributions displayed in Figure \ref{fig:1em4c}). For a probe traveling at $v \sim 0.1 c$, such as the proposed \textit{Breakthrough Starshot}\footnote{https://breakthroughinitiatives.org/initiative/3} initiative, the median of $N_{response}$ is of order unity for $\tau_{obs} = 2.9 \times 10^4 \mathrm{\; yr}$ and the 95\% upper limit of $N_{response}$ is of order unity for $8.1 \times 10^3 \mathrm{\; yr}$, implying that we should not expect a response from a sub-relativistic probe before then (probability distributions displayed in Figure \ref{fig:1em1c}). Figure \ref{fig:ts} illustrates the dependence of the timescale on the speed of the response signal considered. 

\section{Discussion}
\label{sec:d}

For the next few millennia, we should not expect to receive signals or probes from any ETRI on an Earth-like planet orbiting a Sun-like star. In other words, if we receive communication from any such ETI before then, the communicating ETI would most likely have been unaware of humanity's existence as a technological society. 

The number of expected responses to our radio communications is dependent on time since both the distance out to which our radio transmissions reach and the prior for the typical lifetime of an radio-communicating civilization increase with time.

Here, we used the Copernican principle, which represents the most optimistic assumption possible about the prevalence of life-as-we-know-it in the Galaxy, paired with a uniform prior (the most optimistic prior possible), in addition to assuming that all ETIs are ETRIs. Hence, all timescales derived here over which SETRI would be expected to yield negative results are strong lower limits. They should be multiplied by all relevant factors in the Drake equation to yield estimates of when we would expect \textit{positive} results from SETRI.

\vspace*{0.3in} 
\section*{Acknowledgements}
This work was supported in part by the Origins of Life Summer Undergraduate Research Prize Award and a grant from the Breakthrough Prize Foundation.





\bibliography{bib}{}
\bibliographystyle{aasjournal}



\end{document}